\documentclass[
    ,final            
  ]
  {aipproc}

\layoutstyle{6x9}

\begin{document}

\title{Black Hole Accretion: From Quasars to Nano-Quasars}


\author{S. K. Chakrabarti}{
address={S.N. Bose National Center for Basic Sciences, JD-Block, Salt Lake, Kolkata, 700098}
,altaddress={Indian Centre for Space Physics, Chalantika 43, Garia Station Rd., Kolkata, 700084} 
{chakraba@bose.res.in} 
}
\begin{abstract}
In this review we shall comment on a few recent findings which strengthen the view 
that the black hole accretion has substantial amount of sub-Keplerian component. 
The manifestation of this component is many fold. We discuss some of them. A general
outline of the complex structure that emerges from the multitude of observations is
presented. A detailed outline of what might be going on in outburst sources is also
discussed. The relationship amount the spectral and timing properties can be best understood
by this picture. We claim that the sub-Keplerian advective disk paradigm is a complete 
package. Since signatures of sub-Keplerian motion is already increasing in the literature,
the whole package must be correct.
\end{abstract}

\maketitle


\section{Introduction}

The black hole accretion process is very simple in the sense that the flow must be transonic and thus
must cross at least one sonic point. The inner sonic point $r_i$ is located 
between 1 to 6 Kerr radius $r_K=GM/c^2$ (depending on the spin parameter) and 
it is as good as the point of no return as far as the flow is concerned. On the horizon, the 
in-fall velocity is the same as the velocity of light independent of the spin and mass parameter
of the hole.

If the flow has `some' angular momentum ($ \lambda \sim \lambda_{ms}$ to $ \lambda_{mb}$) 
the number of physical sonic points would be more than one and for a significant 
region of the parameter space, the entropy at the inner sonic point $r_i$ is higher 
compared with that at the outer sonic point $r_o$. In these cases, a stable or 
an oscillating shock at a mean radius $r_s$ is allowed at $r_i<r_s<r_o$. The post-shock region 
between the shock and the inner sonic point $r_i$ is known as the CENtrifugal pressure 
dominated BOundary Layer or simply CENBOL where the flow is generally subsonic
and hot. The difference in entropy at the sonic 
points is generated by the shock through turbulent processes in the 
immediate vicinity of the shock. The reason of the shock formation is simple: the centrifugal force
$\lambda^2/r^3$ increases more rapidly than the gravitational force $\sim 1/r^2$ as the flow
comes closer to the hole. This causes the matter to pile up at the centrifugal barrier and form a shock.
For a steady shock, the well known Rankine-Hugoniot conditions, suitably modified to take care of the 
flow model, are satisfied (Chakrabarti, 1989ab). When the shock is non-steady, 
the size of the CENBOL changes rapidly. The CENBOL region is 
the Compton cloud which inverse Comptonizes soft photons (be they from the 
synchrotron emission or from the Keplerian disk, if any). The spectral and 
timing properties of a black hole candidate (whether in a quasar with mass a 
few $\times 10^9 M_\odot$ to a nano-quasar with mass a few $\times M_\odot$) 
is largely governed by the number density of electrons in the cloud (i.e., the 
size and the accretion rate ${\dot M}_h$ of the sub-Keplerian halo) and the 
relative supply of soft-photons from the synchrotron radiation and/or a Keplerian 
disk having a rate of ${\dot M}_d$. 

\section{General Picture from Theoretical Studies}

While the steady disk produces the observed spectrum, the oscillating shock 
produces the quasi-periodic oscillations in the photon counts. There could be 
at least three different ways that the CENBOL can oscillate: (a) when the 
cooling time-scale roughly agrees with the infall time-scale (Molteni,
Sponholz \& Chakrabarti, 1996) (b) when the outflow is important and takes away
matter from the CENBOL (Ryu, Chakrabarti \& Molteni, 1996) and (c) when the CENBOL is at the threshold of 
the optical depth $\tau \sim 1$, i.e., the net accretion rate (${\dot M}_d +{\dot M}_h \sim 1$) (Chakrabarti,
1996, unpublished). In the last case, the CENBOL can oscillate crisscrossing $\tau\sim 1$. This is 
expected to be of high frequency. In general, one can have combinations of 
both (a) and (b) types as seen in Chakrabarti, Acharyya \& Molteni (2004).

Since the inverse of the QPO frequency scales with the mass of the black hole mass, 
in quasars and milli-quasars the frequency would be very low ($\sim$ nano 
to micro Hertz). Furthermore, in these objects the accretion rate is generally 
not high enough to produce (c) type QPOs, thus we do not see high-frequency QPO
equivalent in these objects. The QPOs are likely to have more 
{\it rms} value for higher energy photons since the Comptonized photons 
in CENBOL participate in it. The definition of `high' and `low' energy 
photons, of course, depend on the mass of the black hole. For quasars,
the seed photons are in UV range while for nano-quasars they are
in the soft X-ray range.

As in other compact objects where the boundary layers exist 
the jets are also produced from CENBOL, (Chakrabarti \& Titarchuk, 1995; Chakrabarti et al. 1996; Chakrabarti, 1999).
This is because the Keplerian disks are unable to produce accelerated and collimated jets. The 
outflow rates naturally depend on the shock strength (Chakrabarti, 1999; Das \& Chakrabarti, 1999). 
For a no-shock case, the outflow rate is negligible as in the soft state, while in the very strong shock case,
the outflow rate is low but steady as in the hard states. In the case of intermediate shock
strengths, the outflow rate could be very high and depending on the optical depth of the 
base of the jet below the sonic radius, the jet may or may not be formed. These will be 
akin to burst-on and burst-off states (Chakrabarti \& Nandi, 2000). When the optical depth
of the base of the jet is high enough, there could be return flow bringing matter down 
to the disk, momentarily increasing the local accretion rate. This variation of the accretion rate
causes the variation  of the nature of the light curves as in GRS 1915+105. Here, the 
accretion rate is high enough so that the subsonic part of the jet can crisscross an optical
depth of unity. 

Out of these consistent scenarios, one grand picture of the accretion/outflow process emerges (Fig. 1).
First point is to note that this is far from `self-similar'. In fact, the very cause 
of jet formation is non-self-similarity. Thus, the oft quoted self-similar models of inflow/outflow
are incorrect. The basic components are the Keplerian and the sub-Keplerian 
components, jets/outflows from CENBOL and a return flow from the jet to the disk. However,
their degree of importance will varies: In a very soft state, the CENBOL collapses and the
Keplerian disk reaches close to the black hole, while in a hard state, the CENBOL dominates
and the jets form but not with the highest possible outflow rate.
In the burst-on and burst-off states the jets are also actively participating in the 
variation of light curves. These behaviors should be mass independent, except that in the 
quasars and milli-quasars, the accretion rate may be always sub-Eddington, and thus
variations of the nature of light curve are few and far between.

\section{Origin of the Sub-Keplerian Flow}

A common question which arises in this context is: where does the sub-Keplerian flow 
come from? Independent of whether we have an answer, since all the observational fits necessarily
require them to some degree, they must exist! In the high mass X-ray binaries where there are
profusion of winds, or in quasars and milli-quasars where winds are accreted from multitude 
of surrounding stars, this is not a problem. We do not understand the intermediate-mass black holes
(micro-quasars) to begin with, and so it is not clear what happens there. In the low-mass X-ray binaries
there are evidence of a large Keplerian disk from the amount of time lag and the sub-Keplerian 
component exists but plays a minor role.

By far, the best source of sub-Keplerian flows is the wind from the companion and
perhaps most importantly, ablated matter from the outer region of the Keplerian
disk itself. They subsequently succumb to the gravitational field of the central object. 
Theoretical studies show that the inner edge of the Keplerian disk can be sub-Keplerian 
itself due to variation of transport rate. Even if they originate from 
the Keplerian disk, the infall time is different and that makes an important difference 
in observations (Chakrabarti \& Titarchuk, 1995; Smith et al., 2001, 2002, 2007;
Wu et al., 2002; Soria et al. 2001; Shaposhnikov et al. 2007; Shapohnikov \& Titarchuk, 2006). 

\begin{figure} [h]
\includegraphics[height=13truecm]{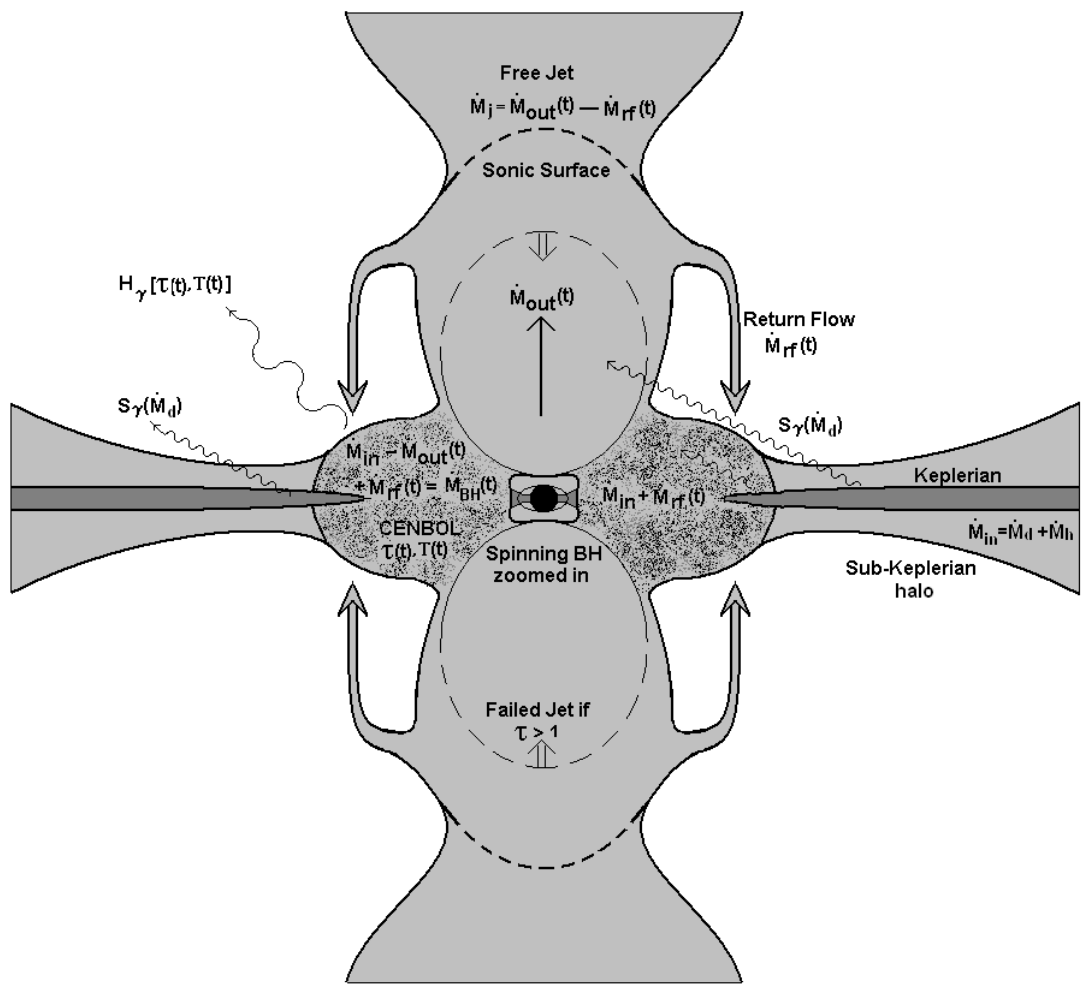}
\caption{A black hole accretion process can be complex with Keplerian and sub-Keplerian flows
accreting simultaneously at ${\dot M}_d$ and ${\dot M}_h$ respectively. 
The outflows and jets may or may not form completely as they depend
on the interaction with the radiation. Each of these components could be non-steady and
could produce a non-steady radiation due to time dependent supply of matter 
at the outer edge of the disk. See text for details.}
\end{figure}

\section{Outburst sources: Need for two components}

It has been shown more than decade ago (Chakrabarti, 1997) that the day to day variation
of the spectrum is often so marked, that it cannot be reconciled with a single component
flow. It was further noted that even if a single Keplerian component becomes 'ADAF' (Narayan \& Yi, 1994)
like flow at the inner part (for hitherto unknown reason) it is difficult to change
the spectral states easily (Chakrabarti, 1997) from hard to soft and vice versa. In Ebisawa et al. 
(1996) a possible scenario of outbursts was presented. With new data 
from a large number of sources over the last decade, this picture requires a refinement.
In Fig. 2, we present the possible scenario which arises out of theoretical works such as
Chakrabarti \& Molteni (1995); Chakrabarti (1996) etc.  After a thorough spectral and temporal 
analysis of several black hole candidates (Chakrabarti et al. 2008; Debnath et al. 2008;
Chakrabarti, Datta \& Pal, 2008; Datta et al. this volume) we can draw a general 
conclusion which, to our knowledge, is unique and has never been reported before. 
We notice that at the rising phase, the shock propagates towards the black hole 
and at the same time, oscillates in a time scale which is proportional to the 
infall time-scale in the CENBOL. Thus the QPO frequency (marked with bi-directional arrows) is found to rise 
monotonically, e.g., in GRS J1655-40, XTE J1550-564 (Chakrabarti et al. 2005, Debnath et al. 
this volume; Datta et al. this volume). The moving in of the shock could either be due to the fact that the 
cooling in CENBOL reduces the post-shock thermal pressure, or because of the excess
ram-pressure of the outbursting inflow in the pre-shock region. The Keplerian disk follows 
the shock and the spectrum becomes softer also. In the phase (d), the shock disappears and 
the Keplerian disk moves in to dominate the outburst. This phase continues as long as
viscosity is high and the Keplerian disk matter is supplied at a high rate. The decline phase
is triggered by the sudden reduction of the Keplerian component, which causes the 
disk to split into two parts. One part moves in towards the black hole, while the {\it inner
edge} of the outer disk recedes. The reduction of ram pressure in the inflow
creates a vacuum  which causes the CENBOL to move {\it outwards}. The QPO frequency
starts to decrease monotonically. The phase (d) need not exist in all the outbursts 
(e.g., 1998 outburst of XTE J1550-564, Datta et al. this volume). In (e) and (f) stages, it may so
happen that the shock starts its journey backward (the flow is still infalling!)
and the Keplerian matter, which lags behind by the viscous time scale, may 
still fall at a high rate towards the black hole. These two components confront in the decline phase,
causing 'hiccups' in the spectral and the temporal properties. The spectral state may become
'intermediate' (neither too soft nor too hard) and QPOs are seen sporadically. Finally as in (g),
the inner Keplerian component is totally accreted and the outer one is receded. Only the shock propagates away
till gets weaker. The spectrum becomes harder and eventually the rate of the sub-Keplerian component 
also starts going down and the object may go to the quiescent state. 

Occasionally in the literature hardness-intensity ratios are plotted for such outbursts
to indicate God's way of writing certain English characters, say `q' (e.g., Belloni, 2006; Homan \& Belloni 2005). 
The problem may arise that since there is nothing special about choosing the specific energy
bands to draw hardness ratio, the diagram is not universal and must change with mass and also
from outburst to outburst. One useful diagram to plot in this context would be to see how the
geometry of the flow changes and this can be done by the taking the ratio of the
photons in the power law (Comptonized photons) and the photons in the disk black body (see photons). 
The ratio would give the degree of interception of the seed photons and thus will contain the
information of the geometry of the flow as time evolves.

\begin{figure} [h]
\includegraphics[height=15truecm]{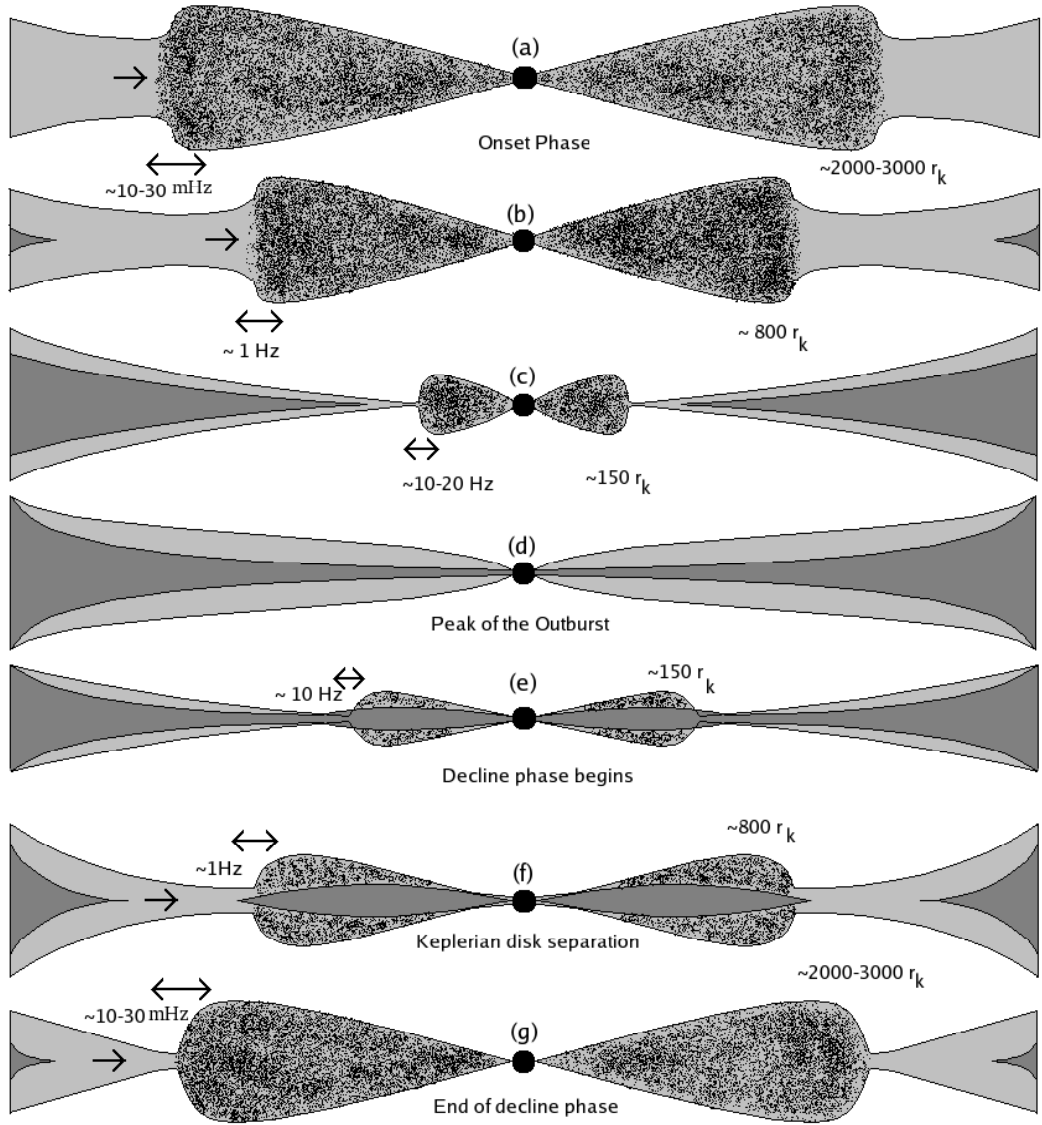}
\caption{A possible scenario of accretion processes in the transient outburst sources from the 
beginning of the rising phase till the end of the decline phase. 
The darker shaded region is the Keplerian disk and lighter shaded region is the sub-Keplerian component.
Dark hazy region is the CENBOL. In the rising phase, the oscillating shock (denoted 
with bi-directional arrows) propagates towards the black hole and in the decline phase it 
propagates backwards. In phases (e) and (f), the shock propagates outward while the 
Keplerian disk continues to move in, causing the intermediate state where QPOs are sporadic. 
Typical values of QPO frequencies and the typical shock locations (in units of Kerr radius $r_K$) is given.}
\end{figure}

\section{Concluding Remarks}

While we do not yet have a complete numerical simulation to show how the
sub-Keplerian matter is supplied to compact objects, we have observations 
to indicate with certainly that they exist. However, a sub-Keplerian flow comes with baggage full of
various ramifications and one has to accept them all. Important among them are the 
existence of standing, oscillating and propagating shocks in the weakly viscous flow. Similarly,
continuous jets and outflows are formed as long as the post-shock region (CENBOL) remains hot (hard state).
Otherwise, the jets could be sporadic and blobby. Shocks can also accelerate particles and synchrotron 
radiations emitted could be extended to very high energies. Fortunately, these are welcome baggages 
as they are exactly what are required to explain the QPOs, the jets and their relation with spectral
state, very high energy radiation etc. Putting all the breads in a single basket comes with a
high risk factor: if there is a strong evidence that the sub-Keplerian component does not exist, 
the whole paradigm collapses! Though it is an either/or situation, we are confident that
sub-Keplerian flow plays a major role in black hole astrophysics. At the same time,
a Keplerian disk, which is fundamentally `sub-sonic' can never have dynamically significant shocks  
(a supersonic to subsonic transition). What is required is a good numerical simulation which 
must include the companion (for nano-quasars) or surrounding (for quasar and milli-quasar) with 
realistic viscous processes to show that pictures presented in Fig. 1 and Fig. 2 are correct. 
Monte-Carlo simulations of Comptonization coupled to hydrodynamic processes are in progress
and the detailed results will be presented elsewhere.

What is obvious from Figs. 1 and 2 is that the magnetic field need not play a major role
in fluid dynamics except perhaps to accelerate and collimate superluminal jets 
(this is hard to do with hydrodynamics alone, though radiative processes can accelerate jets also). It is 
possible that the viscosity is generated by turbulent or magnetic processes (such as Balbus-Hawley instability), 
but it is never too high to eliminate the shocks altogether. 



\end{document}